\documentclass{iacrtrans}

\usepackage{float}
\usepackage{url}
\usepackage{etoolbox}
\makeatletter
\providecommand{\institute}[1]{
  \apptocmd{\@author}{\end{tabular}
    \par
    \begin{tabular}[t]{c}
    #1}{}{}
}
\makeatother

\usepackage{adjustbox}  
\usepackage{placeins}   
\usepackage{longtable}  
\usepackage{xcolor}     
\usepackage{listings}   

\title{Energy Consumption of Post-Quantum Cryptography on Constrained and General-Purpose Architectures}

\author{Olivier Gillot, William J. Buchanan and Madjid G. Tehrani}

\institute{Blockpass ID Lab, Edinburgh Napier University, Edinburgh, UK}

\begin{document}

\maketitle


\begin{abstract}
With Q-day approaching, the transition to post-quantum cryptography (PQC) has begun, with governments across the US, UK and EU mandating migration to quantum-resistant standards. This paper benchmarks the three NIST-standardised PQC algorithms ---FIPS~203 (ML-KEM), FIPS~204 (ML-DSA) and FIPS~205 (SLH-DSA)--- across key generation, signing and verification operations, measuring both computational performance and energy consumption on a range of constrained IoT-class devices, from the 32-bit Raspberry Pi~1 and Zero to the 64-bit Raspberry Pi 4, as well as commodity laptop hardware. Using the FNIRSI FNB58 USB power meter and OpenSSL~3.5, \textcolor{black}{results show that} ML-KEM and ML-DSA achieve energy and speed efficiency comparable to classical elliptic-curve cryptography across all tested architectures. However, SLH-DSA signing is inadvisable on constrained hardware: energy costs for SLH-DSA signing on 32-bit devices were up to 243\% higher than on equivalent 64-bit hardware, making it impractical for resource-limited IoT deployments. These findings have direct implications for IoT security practitioners planning PQC migration.
\end{abstract}

\section{Introduction}
\textcolor{black}{25 years ago, the internet was hardly moderated, and while secure communications have a long history predating the digital era, from wartime cipher machines to early symmetric standards such as DES, encryption for civilian internet traffic was uncommon by default. Anyone could easily access and read all our communications, and there were no limits to the abuses of power that this lack of privacy entailed. However, in time, thanks to the development and large-scale adoption of public-key cryptography --- notably RSA by Rivest, Shamir and Adleman, ECDH by Diffie and Hellman, and AES by Daemen and Rijmen --- cryptography evolved to stronger solutions, with the industry, spearheaded by Google and its ubiquitous Chrome browser, taking the decision to mark any non-encrypted website as insecure, which led to a global implementation of privacy and encryption to much of the internet: currently, between 95 and 99 percent of web navigation in Chrome is protected and inaccessible to eavesdropping thanks to the HTTPS protocol and the TLS standards.~\cite{thompson2025https}}

This shift towards a private and secure internet has thankfully resisted some governments attempts to mandate backdoors~\cite{kleinman2025apple}, however, the encryption methods which have been around for decades are now under a new technological threat: Quantum Computing.~\cite{buchanan2025quantum} By leveraging new progress in experimental physics, scientists have proven that a new type of computer, the Quantum Computer, is so efficient at factorising large numbers that it is estimated that in the near future, using Shor's~\cite{shor1994algorithms} algorithm, public key encryption could be broken. Currently, it only exists on a small scale, but there is a race between state actors to be able to develop one on a large scale, and the date for a fully usable Quantum computer keeps getting closer ~\cite{buchanan2026qday}.

And while AES-256 symmetric encryption seems safe for now~\cite{buchanan2023aes}, since the public key encryption is used precisely to share symmetric keys, \textcolor{black}{it follows that} breaking RSA or ECDH would reveal the private keys, therefore effectively breaking the encryption. Encryption works in such a way that it is either entirely safe or completely broken. The security of an encryption scheme is best understood as a margin that erodes over time: parameters once considered safe weaken as computing power grows, so that much of what was deemed secure in 1980 is now trivial to break. This is what makes ``harvest-now, decrypt-later'' so concerning: an adversary can record today's encrypted traffic and decrypt it once a sufficiently powerful quantum computer exists, exposing communications that were assumed to remain confidential.~\cite{enisa2021current} The answer to this is Post Quantum Cryptography (PQC): a series of algorithms that efficiently protect encryption against hypothetical attacks led by large Quantum computers.

Of course, scientists and governments have been working on solving the challenge. In Europe, the NCSC in Britain~\cite{ncsc2025migrating} and the European Commission~\cite{ec2025pqc} have laid the path to the PQC transition, while the American government agency NIST has launched a vast project to find, select and implement quantum-computing resistant algorithms. At the time of writing, three candidates have been officialised: ML-KEM (Kyber), ML-DSA (CRYSTALS-Dilithium) and SLH-DSA (SPHINCS+)~\cite{csrc2024post}.

Considering the importance of the shift to come, there is a need in the literature for an exploration of the strength and viability of these algorithms. Building on the works of J. Patterson et al.~\cite{patterson2025energy} and Turino et al.~\cite{turino2025pqc}, \textcolor{black}{this paper expands on their ideas} by including signatures and verification beyond the key generation algorithms and to test a variety of machines, including very low powered ones, such as a Raspberry Pi first generation or a Pi Zero, by testing them both on speed and on power consumption. The implementation of the NIST-approved algorithms in OpenSSL~3.5 in April 2025~\cite{openssl2025pqc} \textcolor{black}{provides an opportunity}: it is indeed possible to run OpenSSL on pretty much any device (although installing it does take a few hours on the smaller ones), and \textcolor{black}{it is now possible to compare} both the traditional and PQC algorithms on the same machine without advanced implementation. Hence, by testing variants of the three NIST approved-algorithms in both power consumption and speed against classic cryptography, \textcolor{black}{a broader picture of the viability of PQC on a variety of constrained devices is obtained.} To conduct the tests, \textcolor{black}{a range of Raspberry Pis was used} (32-bit and 64-bit) alongside some ARM and x86 laptops to cover as many architectures as were available \textcolor{black}{for this study.}

\section{Literature Review}
\subsection{Context}
Post-Quantum Cryptography is a very current topic, which has huge implications for both governments and private companies everywhere. There is abundant literature on the topic, but for contextual texts, \textcolor{black}{the reader is referred to} Shor’s foundational paper \cite{shor1994algorithms}, which described in 1994 the theoretical basis on which Quantum computers may be able to break the traditional RSA and Elliptic curve algorithms. The Q-day, this day where quantum computing becomes a reality beyond prototyping, is getting nearer, as underlined by Prof. Buchanan in a recent article \cite{buchanan2026qday}, in which he informs \textcolor{black}{the reader of} the recent advances in qubit research, by Webster \cite{webster2026pinnacle} and Gidney \cite{gidney2025factor}.
It is not only academics who are moving fast on this. The US agency NIST has been proactive in considering the shift towards PQC and has selected three algorithms (with another algorithm, HQC, pending) for standardisation, alongside a deadline for migration for all American Federal systems by 2035. \cite{moody2024transition,moody2025roadahead} In Britain, the NCSC has recommended that a target of 2028 be in place for initial migration, 2031 for complete migration for the highest priority services, and 2035 for complete migration on all systems, services and products \cite{ncsc2025migrating,national2025timelines}. The European Commission has also been proactive on the matter, recommending that all Member States initiate a PQC transition strategy by 2026, that high-risk use cases should be transitioned as soon as possible, no later than 2030 and that by 2035 the transition should be completed as far as possible. \cite{ec2025pqc}
\subsection{Foundational Work}
The main idea behind \textcolor{black}{this paper} is based on the \textit{Energy Consumption Framework and Analysis of Post-Quantum Key-Generation on Embedded Devices} by Patterson \cite{patterson2025energy} Building on it, \textcolor{black}{this study expands on his idea, measuring} PQC energy consumption on a Raspberry Pi device using OpenSSL 3.5 to run the algorithms, and expand on his work by adding a larger variety of devices and by adding signing and verifying alongside key-generation. The reason behind this is that a Raspberry Pi 5 has become quite a capable machine and no longer represents low-power embedded devices as well as an early Raspberry Pi (1 or 2) or a Pi Zero might do. Key-generation alone is not the most significant algorithm for hardware limitations, as the signing results will demonstrate. Finally, and following the example of C.Turino in his paper \textit{PQC-LEO: An Evaluation Framework for Post-Quantum Cryptographic Algorithms} \cite{turino2025pqc}, \textcolor{black}{this paper also tests} algorithms beyond ML-KEM, with signature algorithms ML-DSA and SLH-DSA to explore the actual limitations of smaller hardware, by taking inspiration from his framework.
\subsection{Related PQC Benchmarking Literature}
Tasopoulos et al. \cite{tasopoulos2023energy} measured PQ TLS 1.3 energy consumption on a single STM Nucleo embedded board using WolfSSL. \textcolor{black}{Their measurements, obtained using a high-end oscilloscope, achieved greater precision than the FNB58 USB power meter employed in this study, and encompassed the full TLS 1.3 exchange, which is identified as future work here. The scope of the two studies differs: Tasopoulos et al.\ tested a single purpose-built Cortex-M4 board with a custom WolfSSL implementation, whereas this study covers a range of general-purpose ARM devices running stock OpenSSL 3.5, offering broader hardware coverage and straightforward reproducibility on any Unix machine.} Their work was also published pre-standardisation and therefore does not use the current FIPS names. They found that ``SPHINCS+ needs tremendous amounts of energy and time just for a single handshake.'' \textcolor{black}{a finding consistent with the results reported here for very constrained 32-bit devices, despite the differences in methodology.} 

Paquin et al. \cite{paquin2020benchmarking} benchmarked PQC algorithms in TLS under emulated network conditions, focusing on how packet loss and latency affect performance. \textcolor{black}{The focus on network constraints rather than raw compute power, combined with the use of high-end cloud hardware (Azure D64s with 64 vCPUs) and a custom OQS-OpenSSL fork predating native PQC support in OpenSSL 3.5, makes this work complementary to the present study: the two together address what happens on a constrained device and what happens on a constrained network.} 

Schöffel et al. \cite{schoffel2021energy} evaluated energy costs of PQC KEMs in TLS on a low-power IoT Cortex-M4 device, showing accelerators are not necessary for feasible deployment. \textcolor{black}{The Cortex-M4 at 64 MHz with 256 KB RAM represents a more deeply constrained target than the devices used here, while this study covers a broader range of general-purpose ARM architectures and extends the analysis to signature algorithms ML-DSA and SLH-DSA beyond the KEM focus of Schöffel et al.} Their findings regarding KEMs are consistent with those reported here, and they state that ``proving that the usage of post-quantum KEMs is already feasible today in terms of energy demand and latency for low-power IoT with state-of-the-art hardware'', \textcolor{black}{a conclusion that aligns with the ML-KEM results presented in this work, though SLH-DSA presents a different picture on constrained hardware.} 

S Sarıbaş, S Tonyalı \cite{saribacs2022performance}, Bozhko et al. \cite{bozhko2023performance} measured TLS 1.3 handshake performance (packet overhead and delay) of pre-standardisation PQC algorithms on resource-constrained devices over Ethernet, BLE and WiFi networks. \textcolor{black}{Both studies focus on network-layer metrics --- packet overhead and handshake delay --- rather than local energy consumption, and neither includes power measurements. This makes them complementary to the present study, which focuses on per-device energy and timing. The use of a Raspberry Pi 4 as the network endpoint in Bozhko et al.\ is noted; as the results here illustrate, the Pi 4 is a capable machine and less representative of deeply constrained hardware than earlier Pi generations.} Network and packet capture would fall neatly into a future expansion of this work. 

Roma et al. \cite{roma2021energy} analysed the energy consumption of all NIST Round 2 PQC candidates on a single Intel i7 using software-based power measurement via PAPI/RAPL. \textcolor{black}{The use of software-based estimation rather than physical measurement is a deliberate methodological difference: software power measurement is acknowledged to be imprecise, as Roma et al.\ themselves note --- ``which makes it difficult to make a correlation between the energy measurements obtained in this work and the energy consumption expected on a low-end embedded processor.'' The present study addresses this gap through physical inline measurement on low-end ARM hardware, covering the standardised NIST algorithms against classical baselines, making the two works complementary in scope.} 

In a recent paper (January 2026), Algar-Fernandez et al. \cite{algar2026benchmarking} benchmarked the timing performance of standardised and research-stage PQC primitives across two commodity Windows CPUs using a Python TCP testbed without energy measurements. \textcolor{black}{The study focuses on execution time across a limited set of x86 hardware configurations and, as the authors acknowledge, does not cover energy consumption, memory usage, or the diversity of deployment environments including microcontrollers and Linux-based embedded architectures. The present study addresses these dimensions across a range of ARM devices and x86 laptops running OpenSSL 3.5 on Linux and macOS, offering a complementary perspective on deployment-relevant metrics.}

\section{Implementation}
\subsection{Hardware}
\subsubsection{The Pis}
\textcolor{black}{Following the literature review,} the focus was placed on Raspberry Pi from the early generations and smaller processors. \textcolor{black}{Four devices were already available from past projects,} requiring only the procurement of a Raspberry Pi 2 v1.1, Raspberry Pi 2 v1.2 (whilst very similar, the second one is actually 64-bit enabled, leading to an interesting test of the practical difference between 32 and 64-bit architecture on ARM devices) and a Raspberry Pi 3 (Model A+).

The Raspberry Pis do not have TPMs (Trusted Platform Modules) implemented natively. It exists as a separate module \cite{pihut2026tpm} but has not been tested in this study, and all cryptographic operations are done directly at the software level.

This \textcolor{black}{resulted in} a lineup of seven different ARM machines, ranging from a 700 MHz single-core and 512 MB of memory for the Raspberry Pi 1 to the 1.5 GHz quad-core Pi 4 with 8 GB of LPDDR4 memory, a very capable machine.

\begin{table}
\centering
\caption{Raspberry Pi Hardware Specifications}
\label{tab:raspberry-pi-specs}
\begin{adjustbox}{max width=\textwidth}
\begin{tabular}{lllll}
\hline
\textbf{Model} & \textbf{OS} & \textbf{CPU} & \textbf{Memory} & \textbf{Arch} \\
\hline
Raspberry Pi 1 Model B & Debian Bookworm & 700 MHz single-core ARM1176 & 512 MB LPDDR2 & 32-bit (ARM) \\
\hline
Raspberry Pi 2 Model B v1.1 & Debian Bookworm & 900 MHz quad-core ARM Cortex-A7 (ARMv7) & 1 GB SDRAM (450 MHz) & 32-bit (ARM) \\
\hline
Raspberry Pi 2 Model B v1.2 & Debian Bookworm & 900 MHz quad-core ARM Cortex-A53 (ARMv8) & 1 GB SDRAM (450 MHz) & 64-bit (ARM) \\
\hline
Raspberry Pi Zero WH & Debian Bookworm & 1 GHz single-core ARM1176 & 512 MB LPDDR2 & 32-bit (ARM) \\
\hline
Raspberry Pi Zero 2 WH & Debian Bookworm & 1 GHz quad-core ARM Cortex-A53 & 512 MB LPDDR2 & 64-bit (ARM) \\
\hline
Raspberry Pi 3A+ & Debian Bookworm & 1.4 GHz quad-core ARM Cortex-A53 & 512 MB LPDDR2 & 64-bit (ARM) \\
\hline
Raspberry Pi 4 & Debian Bookworm & 1.5 GHz quad-core Cortex-A72 & 8 GB LPDDR4 & 64-bit (ARM) \\
\hline
\end{tabular}
\end{adjustbox}
\end{table}

Figure~1 presents all the Raspberry Pis and the FNB58 used in the experiments.

\begin{figure}
    \centering
    \includegraphics[width=1\linewidth]{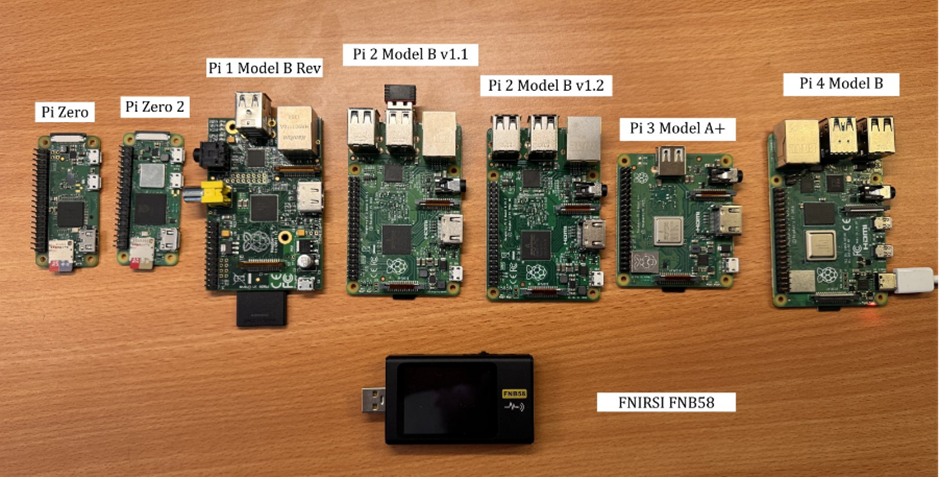}
    \caption{Raspberry Pi lineup and FNB58}
    \label{fig:labsetup}
\end{figure}

\subsubsection{The FNIRSI FNB58}
After researching it on specialised websites \cite{elektor2023fnb58}, \textcolor{black}{the FNIRSI FNB58 was selected as the power measurement tool for this study. Its simplicity of use, accurate readings and reasonable price made it the ideal choice.} It is an inline USB power meter which measures voltage and current continuously. \textcolor{black}{Power (PBUS) in watts was collected} at 50 samples per second, logged via the proprietary USB-meter software.

\begin{figure}
    \centering
    \includegraphics[width=1\linewidth]{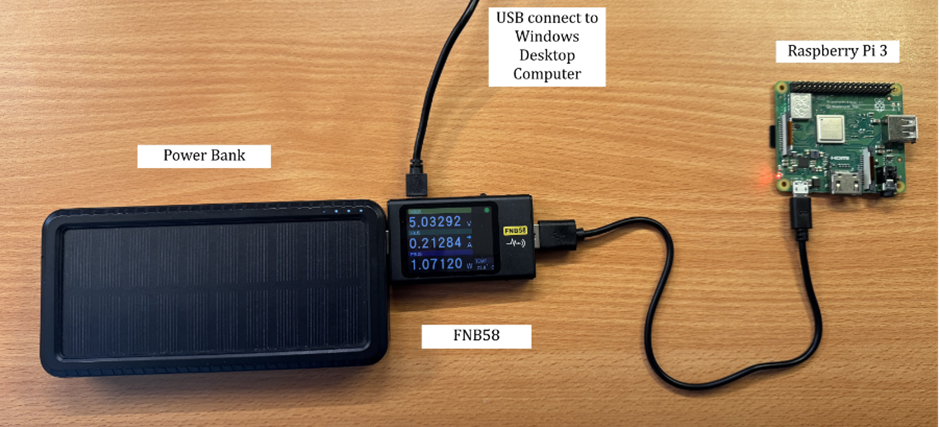}
    \caption{FNB58 in operation}
    \label{fig:fnb58}
\end{figure}

\textcolor{black}{Operation required connection between} a power bank for 5 V (up to 2.1 A) power delivery and a micro-USB cable connected to the Raspberry Pi to power it and record the power variations. Figure~2 presents the FNB58 in operation.

\subsubsection{Laptops and Desktop Machines}
\textcolor{black}{Three laptops were also included:} an Ubuntu Thinkpad T480s Intel from 2017, a MacBookAir M1 (ARM) from 2020 and a 2024 Thinkpad P14s Gen5 AMD. Due to the complexity of isolating power readings for the OpenSSL load and the unreliability of software estimations, \textcolor{black}{only timed tests were conducted on these machines.}.

\begin{table}
\centering
\caption{Laptop Hardware Specifications}
\label{tab:laptop-specs}
\begin{adjustbox}{max width=\textwidth}
\begin{tabular}{llll}
\hline
 & \textbf{Thinkpad T480s Intel} & \textbf{MacBookAir M1} & \textbf{Thinkpad P14s Gen5 AMD} \\
\hline
\textbf{OS} & Ubuntu 25.10 & macOS Tahoe 26.3 & Ubuntu 24.04 \\
\hline
\textbf{CPU} & Intel Core i7-8550U & Apple M1 (ARM) & AMD Ryzen 5 PRO 8640HS \\
\hline
\textbf{Memory} & 40  GB DDR4 & 8  GB LPDDR4X & 16  GB DDR5 \\
\hline
\textbf{Arch} & 64-bit (x86-64 Intel) & 64-bit (ARM-M1) & 64-bit (x86-64 AMD) \\
\hline
\end{tabular}
\end{adjustbox}
\end{table}

\textcolor{black}{The} Windows 11 AMD Desktop machine was used to coordinate and control all the different machines via WSL Ubuntu and SSH, and to gather the FNB58 data via a micro-USB cable on their proprietary Windows-only software.

\subsection{Software}
\subsubsection{OSes}
Each of the Raspberry Pi machines was freshly imaged using the Raspberry Pi Imager utility to flash a lightweight desktop-free Debian Bookworm distro on a micro-SD card (and SD card for the Raspberry Pi 1). This ensured that the very minimum of other processes would interfere with \textcolor{black}{power reading collection.}

Both Thinkpad Laptops were running Ubuntu, and the MacBook Air runs macOS Tahoe. Having all machines running on Unix \textcolor{black}{allows the same programs to be run} without having to have multiple versions. Since \textcolor{black}{power readings were not collected on these machines,} fresh installs were not required\textcolor{black}{; all irrelevant programs and processes were disabled when speed tests were conducted.}

\subsubsection{OpenSSL 3.5}
Fundamental to this paper, the stock version of OpenSSL 3.5 includes all the PQC algorithms \textcolor{black}{central to this study's testing} (ML-KEM, ML-DSA, SLH-DSA) \cite{openssl2025pqc}. By being able to run exactly the same program (OpenSSL) on a variety of algorithms on a variety of machines \textcolor{black}{ensures fair readings are obtained,} depending only on variations of hardware instead of different software implementations.

Installing OpenSSL proved to be a rather lengthy process for the weaker Pis. It took several hours to install it on the Raspberry Pi 1 and the Pi Zero; however, \textcolor{black}{as the results demonstrate,} it is doable. Whilst OpenSSL ships as standard on Linux distributions and MacOS, \textcolor{black}{the 3.5 version was required,} which was not included in any of the distros used, therefore requiring a manual build from source for all the machines. All the command lines used are to be found in the appendix and on the GitHub of this project.

\paragraph{Key Generation}
All algorithms tested \textcolor{black}{in this study} generate a private/public key pair.
For all PQC and classical algorithms except RSA, the private key is generated with:
\begin{lstlisting}[language=bash]
openssl genpkey -algorithm [ALGORITHM] -out private.key
\end{lstlisting}
RSA uses:
\begin{lstlisting}[language=bash]
openssl genrsa -out private.key 2048
\end{lstlisting}
The public key is then derived for all algorithms with:
\begin{lstlisting}[language=bash]
openssl pkey -in private.key -pubout -out public.key
\end{lstlisting}
The keys are fundamental to all cryptographic operations. ML-KEM keys are used for encapsulation and decapsulation, operations indispensable for TLS; full TLS certificate generation and handshake are left as future work.

\paragraph{Signing}
For ML-DSA and SLH-DSA, the private key is used to sign data and the public key to verify it. \textcolor{black}{In this study}, the \texttt{README.md} file of the repository (19.6\,KB) was used as the message to be signed and verified. A single key pair is generated with the commands above, then:
\begin{lstlisting}[language=bash]
openssl pkeyutl -sign -inkey private.key -in README.md \ 
    -out signature.bin
\end{lstlisting}

\paragraph{Verification}
The signature is verified using the public key generated alongside the private key:
\begin{lstlisting}[language=bash]
openssl pkeyutl -verify -pubin -inkey public.key \
    -in README.md -sigfile signature.bin
\end{lstlisting}

\begin{table}
\centering
\caption{Algorithm key and signature sizes. PQC signature sizes verified from repository data; classical and key sizes from FIPS~203, 204, 205.}
\label{tab:algo_sizes}
\resizebox{\textwidth}{!}{%
\begin{tabular}{llcrrrc}
\hline
\textbf{Algorithm} & \textbf{Type} & \textbf{FIPS} & \textbf{Pub Key (B)} & \textbf{Priv Key (B)} & \textbf{Signature (B)} & \textbf{NIST Level} \\
\hline
RSA-2048             & Sign/KEM  & ---  & 256   & 256   & 256    & ---  \\
ECDSA P-256          & Signature & ---  & 64    & 32    & 64     & ---  \\
ECDH P-256           & KEM       & ---  & 64    & 32    & N/A    & ---  \\
X25519               & KEM       & ---  & 32    & 32    & N/A    & ---  \\
ML-KEM-512           & KEM       & 203  & 800   & 1632  & N/A    & 1    \\
ML-KEM-768           & KEM       & 203  & 1184  & 2400  & N/A    & 3    \\
ML-KEM-1024          & KEM       & 203  & 1568  & 3168  & N/A    & 5    \\
ML-DSA-44            & Signature & 204  & 1312  & 2528  & 2420   & 2    \\
ML-DSA-65            & Signature & 204  & 1952  & 4000  & 3309   & 3    \\
ML-DSA-87            & Signature & 204  & 2592  & 4864  & 4627   & 5    \\
SLH-DSA-SHA2-128s    & Signature & 205  & 32    & 64    & 7856   & 1    \\
SLH-DSA-SHA2-128f    & Signature & 205  & 32    & 64    & 17088  & 1    \\
SLH-DSA-SHAKE-128s   & Signature & 205  & 32    & 64    & 7856   & 1    \\
SLH-DSA-SHA2-192s    & Signature & 205  & 48    & 96    & 16224  & 3    \\
SLH-DSA-SHA2-192f    & Signature & 205  & 48    & 96    & 35664  & 3    \\
SLH-DSA-SHAKE-192s   & Signature & 205  & 48    & 96    & 16224  & 3    \\
SLH-DSA-SHA2-256s    & Signature & 205  & 64    & 128   & 29792  & 5    \\
SLH-DSA-SHA2-256f    & Signature & 205  & 64    & 128   & 49856  & 5    \\
SLH-DSA-SHAKE-256s   & Signature & 205  & 64    & 128   & 29792  & 5    \\
\hline
\end{tabular}%
}
\end{table}

\subsubsection{USB-meter and .cfn format conversion}
To record the power readings, another fundamental point of the project, \textcolor{black}{power readings were recorded using} the proprietary program, the USB-meter v0.7, found on the FNIRSI website. \cite{fnirsi2025software} \textcolor{black}{It enabled viewing and recording of} the PBUS (watts) data, the most important for \textcolor{black}{this study's} data collection, as seen in Figure~3:

\begin{figure}
    \centering
    \includegraphics[width=1\linewidth]{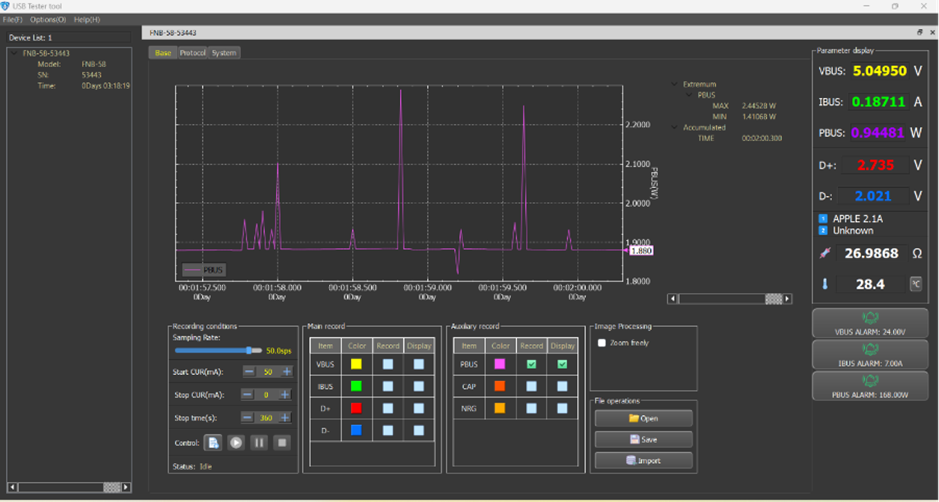}
    \caption{USB-meter in operation}
    \label{fig:usb-meter}
\end{figure}

The program shipped with the FNB58 unfortunately only saves data with a proprietary format, .cfn. To circumvent the problem, \textcolor{black}{a utility was used} to turn these files into .csv files. \cite{volkov2025usbmeter}

\subsection{Measurement Procedures and Data Collection}
\subsubsection{Power Readings}
Once all the hardware had been procured and the software installed, the first step was to gather accurate power readings for each of the seven Raspberry Pis.

\textcolor{black}{Each machine followed the same steps. After a few minutes of warm-up, a connection was established to the Pi via SSH to launch the commands. A check was always performed to ensure} no processes were running in the background, and then, from the USB-meter utility, \textcolor{black}{a} five-minute baseline at 50 samples per second PBUS \textcolor{black}{was recorded} for idle. \textcolor{black}{The file was saved as a .cfn file, then converted into a usable .csv file using usb-meter-utils.} It is then simple to average the PBUS column values to get an average baseline idle power reading.

\textcolor{black}{pqc\_benchmark.py (described below) was then used} to select which method or algorithm to run, \textcolor{black}{with} a very large number of iterations \textcolor{black}{selected} to make sure it does not finish before the recording is done, and \textcolor{black}{recordings taken} at approximately 30-second intervals whilst the program is on-load. This relatively small interval (which collected circa 1,5000 samples at 50 samples per second) was due to time constraints: having to run 47 algorithms per machine, for seven machines, led to 329 samples to take. \textcolor{black}{The same procedure was then followed} as for the baseline to collect the average of on-load power readings. All the .cfn and .csv files are available for review on the GitHub.

Energy per iteration is calculated using the delta power method. The average on-load power is subtracted from the average baseline idle power to isolate the cryptographic operation's contribution:

\begin{equation}
{E_{op} = (P_{load} - P_{idle}) \times T_{op}}
\end{equation}

{where $P_{load}$ is the average on-load power (W), $P_{idle}$ is the average baseline idle power (W), $T_{op}$ is the mean time per iteration (s), and $E_{op}$ is the energy per iteration (J).}

Tables summing up the power readings are available in the Appendix.

\subsubsection{Python Wrappers Programs}

\textcolor{black}{Python programs were created} as wrappers for the OpenSSL operations. Both programs (pqc\_benchmark.py and pqc\_benchmark\_automated.py) are available for review on the GitHub of the paper.

They simply compile all the OpenSSL commands, including creating keys, creating certificates, creating a message to be signed and/or verified, for every algorithm tested in a convenient program which lets the user select the number of iterations, and offers the possibility of an automated option for fast benchmarking. The power values may be optionally added, and all results are given a csv file.

Anyone with OpenSSL 3.5 enabled (complete steps for building it are in the GitHub of the project) running on a Unix machine can run both these programs. Please note that both programs are needed to run the automated benchmark, and that multiple files -keys, certificates, message to be signed and verified) are created in the process in a sub-folder.

\subsubsection{Time Metrics Readings}

To get the time metrics, two options are available: a fast, automated test where the Python wrapper program only takes one variable, the number of iterations, and automatically runs every available algorithm (pqc\_benchmark\_automated.py) and puts them in a CSV results file. This makes it ideal for testing any Unix machine without power readings involved. The second option has more granularity, with a Python wrapper which lets the user select the method, the algorithm, and enter the power metrics to increment a CSV results file.

\subsubsection{Running the Tests}

\textcolor{black}{During data collection, every} algorithm and method \textcolor{black}{was run} through the power metrics first, gathering the baselines and on-load power readings for every machine, method and algorithm for each of the seven Raspberry Pis.
\textcolor{black}{Each method was then run} with each algorithm on each machine, providing the gathered power readings, on 100 iterations (except when the extreme slowness of SLH-DSA on slow devices \textcolor{black}{limited this to} 10) to gather \textcolor{black}{the} final results, samples of which are in the appendix and can be seen on GitHub. Below is Table~3 containing a small sample of the results, presenting the number of Joules needed per iteration.

\begin{table}
\centering
\scriptsize
\caption{Key generation and signing, Joules per iteration across Raspberry Pi models}
\begin{tabular}{lccccccc}
\hline
 & pi1 & pizero & pi2 & pi2b & pizero2 & pi3 & pi4 \\
\hline

\multicolumn{8}{l}{\textbf{Keys}} \\

ecdh & 0.0483 & 0.0500 & 0.0380 & 0.0346 & 0.0382 & 0.0436 & 0.0319 \\
ml-kem-512 & 0.0519 & 0.0517 & 0.0398 & 0.0357 & 0.0388 & 0.0428 & 0.0345 \\
ml-dsa-44 & 0.0525 & 0.0519 & 0.0412 & 0.0361 & 0.0371 & 0.0444 & 0.0347 \\
slh-dsa-sha2-128f & 0.0601 & 0.0644 & 0.0503 & 0.0431 & 0.0464 & 0.0525 & 0.0407 \\
slh-dsa-sha2-128s & 0.7259 & 0.9182 & 0.7093 & 0.5521 & 0.6265 & 0.6304 & 0.4612 \\
slh-dsa-shake-192s & 2.0534 & 2.6092 & 2.2992 & 0.6654 & 0.7613 & 0.8485 & 0.6370 \\

\hline
\multicolumn{8}{l}{\textbf{Signing}} \\

ecdsa & 0.0254 & 0.0258 & 0.0206 & 0.0184 & 0.0192 & 0.0241 & 0.0197 \\
ml-dsa-44 & 0.0303 & 0.0320 & 0.0251 & 0.0218 & 0.0214 & 0.0284 & 0.0200 \\
slh-dsa-sha2-128f & 0.2757 & 0.3776 & 0.2711 & 0.2103 & 0.2359 & 0.2396 & 0.1730 \\
slh-dsa-sha2-128s & 5.1354 & 7.2222 & 5.1381 & 3.9943 & 4.5117 & 4.5984 & 3.2258 \\
slh-dsa-shake-192s & 18.7602 & 24.7630 & 19.6558 & 5.7308 & 6.5288 & 7.3749 & 5.4714 \\

\hline
\end{tabular}
\end{table}

\section{Evaluation}
Tables containing a sample of the test results can be found in the Appendix.
\subsection{Viability of ML-KEM on Constrained Hardware Compared to ECDH and X25519 for Key-Generation}

As Figure~4 shows, the ML-KEM (Kyber) FIPS 203 algorithm holds its ground when it comes to key-generation compared to the classic ECDH and X25519 elliptic curves. ML-KEM key generation remained within 2-12\% of ECDH and X25519 across all tested devices, with the single exception of ML-KEM-1024 on the Pi1 at 17\% overhead and the noticeably good results of Pizero 2 being actually marginally faster than ECDH. Schöffel et al. \cite{schoffel2021energy} specifically concluded: ``the usage of post-quantum KEMs is already feasible today in terms of energy demand and latency for low-power IoT'', \textcolor{black}{consistent with the findings of this study, which confirm that} the ML-KEM algorithms are a perfectly viable Quantum-safe replacement for the currently used but vulnerable elliptic curves.

It performed across the board, and even on very constrained 32-bit hardware such as the Pi 1 or the Pi Zero, making it perfectly plausible on a range of IoT devices.

\begin{figure}
    \centering
    \includegraphics[width=1\linewidth]{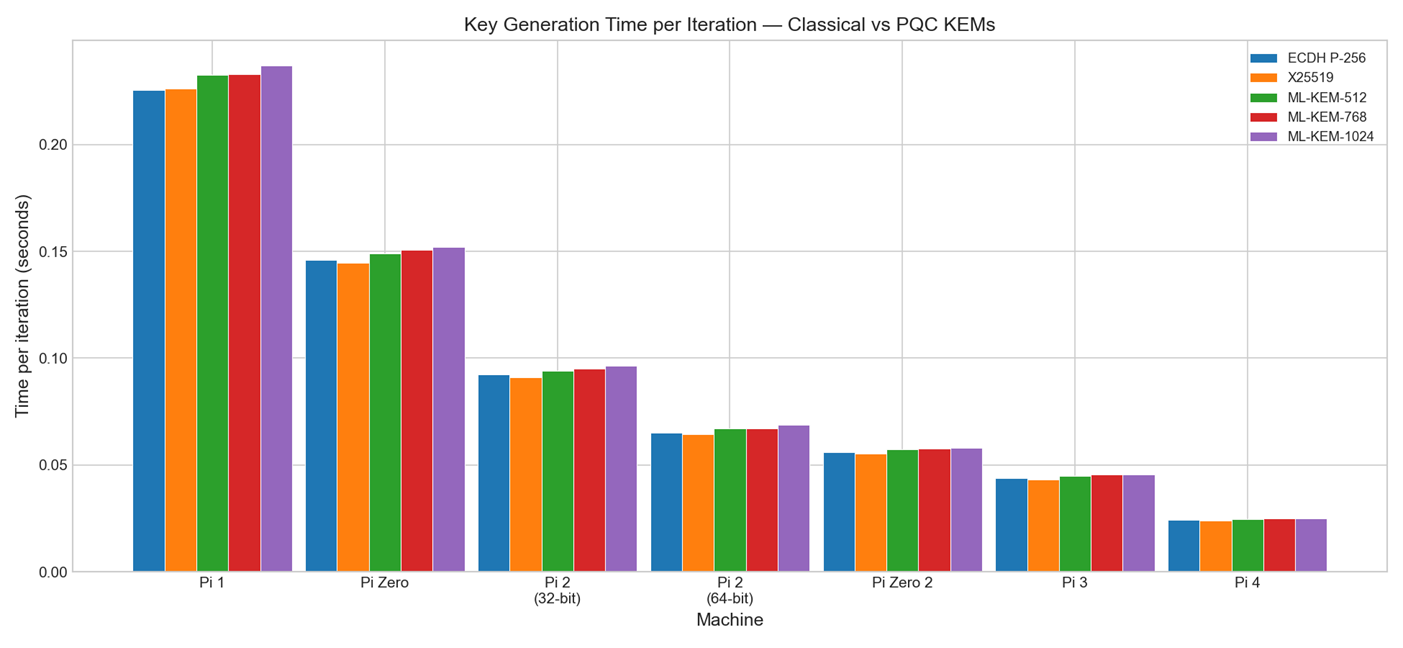}
    \caption{ML-KEM on constrained hardware compared to ECDH and X25519}
    \label{fig:kengen-mlkem}
\end{figure}

\subsection{SLH-DSA vs ML-DSA Performance and Energy Costs for Signatures}

\begin{figure}
    \centering
    \includegraphics[width=1\linewidth]{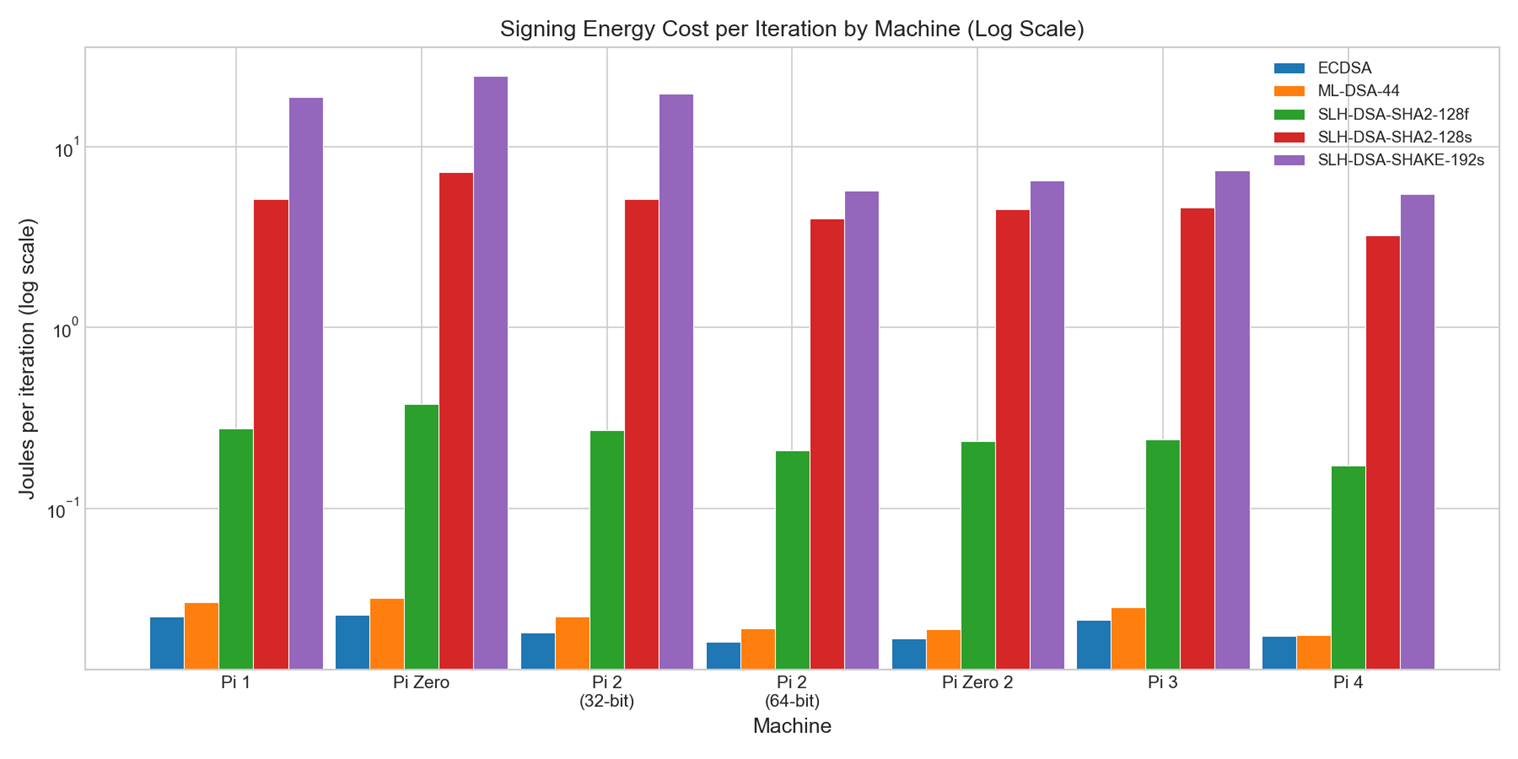}
    \caption{SLH-DSA vs ML-DSA signing energy consumption (Joules) across Raspberry Pi models, logarithmic scale}
    \label{fig:slh-vs-mldsa}
\end{figure}

As shown in Figure~5, whilst ML-DSA-44 (Dilithium) FIPS 204 got near the performance of ECDSA for signing, with a difference diminishing as the hardware gets more performant, from 19-24\% more energy used than ECDSA for the 32-bit devices to a 1.3\% on the Pi 4, it paints a very different picture for the SLH-DSA (SPHINCS+) FIPS 205 algorithms. \textcolor{black}{A} logarithmic scale \textcolor{black}{was} needed to \textcolor{black}{illustrate} how quickly the signature costs in Joules increase exponentially with these intentionally slow algorithms. They were designed to prioritise security conservatism through complex Merkle trees over performance, never to be used extensively and repeatedly on slow hardware as emphasised by Tasopoulos et al. ``the NIST conservative security choice of SPHINCS+ as a Digital Signature (…) can only be used in a unilateral authentication since SPHINCS+ needs tremendous amounts of energy and time just for a single handshake'' \cite{tasopoulos2023energy}, and \textcolor{black}{the results presented here clearly match theirs and illustrate why.} It is interesting to note that whilst it is possible to sign with SLH-DSA-SHAKE-192s on a Pi 1, it comes at such a cost that it may not be advisable, as the next section will show.

\subsection{Energy vs. Speed: Key-Generation ML-KEM-768 and Signing SLH-DSA-SHAKE-192s}

\textcolor{black}{Figure~6 presents} the costs in Joules of 100 Key-Generation with ML-KEM 768 alongside how a very low power draw on the smaller CPU machines and 32-bit architecture nevertheless ends up consuming more Joules for Key-Generation. The reason is simple, whilst using a lot less power ---the Pi 1 on-load delta draw is 0.212 W, 6.6 times less than the Pi 4 1.41 W delta draw for the same operation!--- the speed of execution is the deciding factor, and the Pi 4, despite drawing more power, beats every other machine by simply being the fastest.

\begin{figure}
    \centering
    \includegraphics[width=1\linewidth]{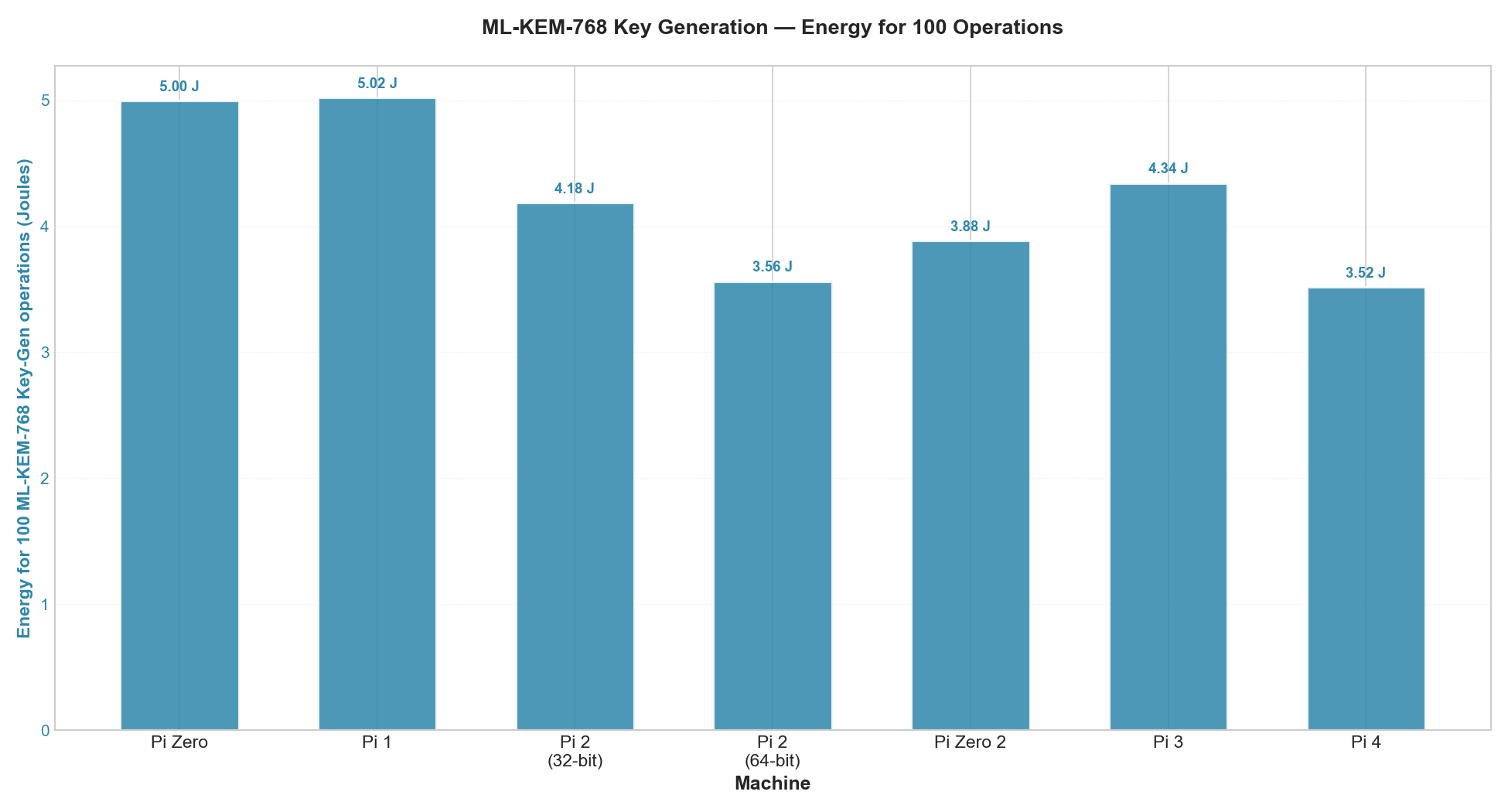}
    \caption{ML-KEM-768 key generation energy consumption (Joules) across Raspberry Pi models}
    \label{fig:mlkem-energy}
\end{figure}

The same comparison applies to SLH-DSA, and it reveals that the decisive factor is the algorithm rather than the operation. For ML-KEM-768 key generation, the 32-bit Pi2 v1.1 needs only about 17\% more energy than the 64-bit Pi2 v1.2. For SLH-DSA-SHAKE-192s, however, the 32-bit penalty is severe in both operations: roughly 245\% more energy for key generation (2.2992 J vs 0.6654 J) and 243\% more for signing (19.6558 J vs 5.7308 J). It is the hash-heavy design of SLH-DSA, not the choice of operation, that makes it hostile to 32-bit hardware. Whilst ML-KEM remains a perfectly viable option on slow devices, SLH-DSA is impractical on 32-bit architectures regardless of whether it is generating keys or signing.

\begin{figure}
    \centering
    \includegraphics[width=1\linewidth]{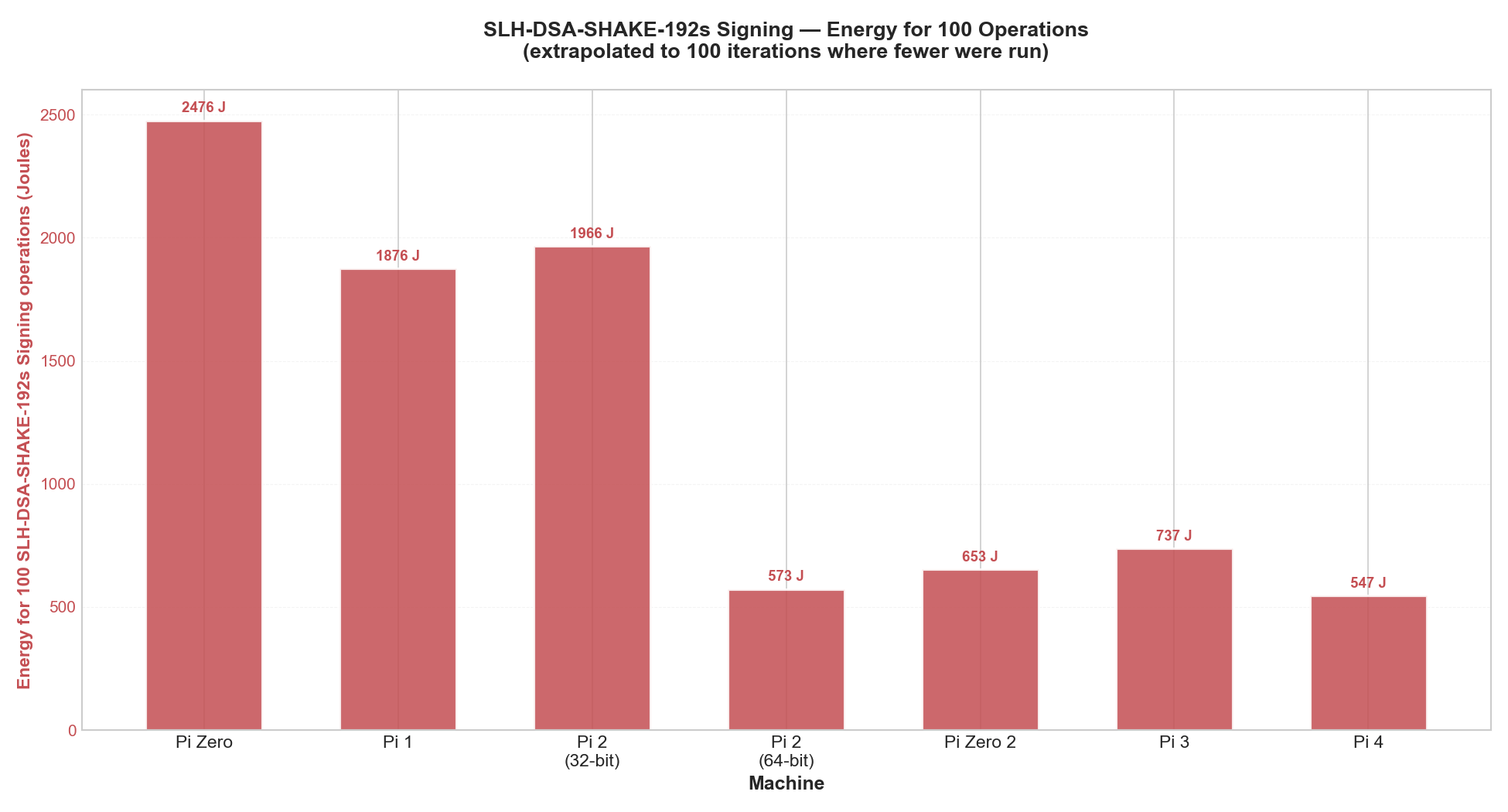}
    \caption{SLH-DSA-SHAKE-192s signing energy consumption (Joules) across Raspberry Pi models}
    \label{fig:slhsign}
\end{figure}

\subsection{Timing Performance on x86 and ARM Laptop vs Pi4}

\textcolor{black}{A limitation of this study was the inability} to gather accurate power readings for the tests running on the different laptops \textcolor{black}{included in the study.} However, the timing metrics were easy to gather and helped serve as a point of reference against the Pis.
\textcolor{black}{As shown in Table~5,} the modern AMD Thinkpad is the fastest overall, only beaten by the M1 on SLH-DSA-SHAKE-192s signing, and The MacBook Pro M4 figures are for completeness, not directly comparable and should not be read as a like-for-like measurement: they were collected on a different OpenSSL build (3.6), with verification omitted, and the resulting times are implausibly slow for the hardware (e.g. ECDSA signing at 0.0377 s versus 0.0064 s on the M1). \textcolor{black}{It is also notable that} the Pi4 clearly holds its own respectably against much more capable machines on ML-KEM and ML-DSA, blurring the line between a laptop and what is perhaps inappropriately considered an embedded or IoT-class device.

\begin{table}
\centering
\scriptsize
\caption{Cryptographic Performance --- Time per Iteration (s) across laptops and Raspberry Pi 4. M4 data courtesy of Bill Buchanan.}
\begin{adjustbox}{max width=\textwidth}
\begin{tabular}{llccccc}
\hline
Algorithm & Operation & Pi 4 & MacBook Air M1 & ThinkPad Intel & ThinkPad AMD & MacBook Pro M4 \\
\hline

\multicolumn{7}{l}{\textbf{Key Generation}} \\

RSA-2048 & KeyGen & 0.3526 & 0.0576 & 0.0538 & 0.0408 & 0.0868 \\
ECDSA P-256 & KeyGen & 0.0247 & 0.0124 & 0.0104 & 0.0072 & 0.0333 \\
X25519 & KeyGen & 0.0239 & 0.0118 & 0.0111 & 0.0073 & 0.0328 \\
ECDH P-256 & KeyGen & 0.0243 & 0.0123 & 0.0114 & 0.0074 & 0.0333 \\
ML-KEM-512 & KeyGen & 0.0248 & 0.0123 & 0.0108 & 0.0077 & 0.0341 \\
ML-DSA-44 & KeyGen & 0.0255 & 0.0125 & 0.0099 & 0.0078 & 0.0324 \\
SLH-DSA-SHA2-128f & KeyGen & 0.0297 & 0.0127 & 0.0142 & 0.0080 & 0.0352 \\
SLH-DSA-SHA2-128s & KeyGen & 0.3241 & 0.0390 & 0.1177 & 0.0330 & 0.1270 \\
SLH-DSA-SHAKE-192s & KeyGen & 0.5114 & 0.1200 & 0.1885 & 0.1161 & 0.1487 \\

\hline
\multicolumn{7}{l}{\textbf{Signing}} \\

RSA-2048 & Sign & 0.0194 & 0.0072 & 0.0049 & 0.0033 & 0.0199 \\
ECDSA P-256 & Sign & 0.0136 & 0.0064 & 0.0051 & 0.0036 & 0.0377 \\
ML-DSA-44 & Sign & 0.0156 & 0.0069 & 0.0074 & 0.0047 & 0.0305 \\
SLH-DSA-SHA2-128f & Sign & 0.1234 & 0.0162 & 0.0450 & 0.0129 & 0.0518 \\
SLH-DSA-SHA2-128s & Sign & 2.2925 & 0.2096 & 0.8240 & 0.1945 & 0.7278 \\
SLH-DSA-SHAKE-192s & Sign & 4.4086 & 0.9783 & 1.6120 & 0.9871 & 1.0484 \\

\hline
\multicolumn{7}{l}{\textbf{Verification}} \\

RSA-2048 & Verify & 0.0101 & 0.0059 & 0.0045 & 0.0033 & N/A \\
ECDSA P-256 & Verify & 0.0112 & 0.0060 & 0.0057 & 0.0033 & N/A \\
ML-DSA-44 & Verify & 0.0124 & 0.0061 & 0.0047 & 0.0037 & N/A \\
SLH-DSA-SHA2-128f & Verify & 0.0170 & 0.0066 & 0.0085 & 0.0040 & N/A \\
SLH-DSA-SHA2-128s & Verify & 0.0125 & 0.0062 & 0.0063 & 0.0038 & N/A \\
SLH-DSA-SHAKE-192s & Verify & 0.0138 & 0.0068 & 0.0065 & 0.0042 & N/A \\

\hline
\end{tabular}
\end{adjustbox}
\end{table}

\subsection{Summary of Findings}

Post-quantum algorithms ML-KEM and ML-DSA demonstrate viability for embedded systems, with performance moderately degraded on the slowest Pis (Pi1 and Pi zero) and only marginally degraded on the faster machines compared to classical Elliptic Curve algorithms. However, SLH-DSA, while providing maximum security, is computationally impractical for real-time applications, particularly on resource-constrained devices. This analysis demonstrates that lattice-based PQC (ML-KEM/ML-DSA) represents the viable path forward for IoT and embedded quantum-safe cryptography.

\section{Conclusions}
This study evaluated the viability of the three NIST-standardised PQC algorithms, FIPS 203 (ML-KEM), FIPS 204 (ML-DSA) and FIPS 205 (SLH-DSA), on constrained IoT-class devices ranging from the 32-bit Pi 1 and Pi Zero to the 64-bit Pi 4, testing key generation, signing and verification against elliptic curve baselines using stock OpenSSL 3.5 and physical energy measurement via the FNIRSI FNB58.
\textcolor{black}{The results show that,} alongside what had already been found by Patterson \cite{patterson2025energy} ---whose work inspired \textcolor{black}{this study}--- and Schöffel \cite{schoffel2021energy}, ML-KEMs and ML-DSAs are ready for deployment, on any type of device and architecture tested. They consistently matched or closely matched Elliptic Curves in energy consumption and speed tests. However, and much like Tasopoulos \cite{tasopoulos2023energy} stated, SLH-DSA, particularly for signing, is not suitable for constrained devices, with a particular emphasis on a 243\% overhead in energy consumption uniquely based on the difference between a 32-bit and 64-bit architecture; therefore, 32-bit devices should not be considered for SLH-DSA deployment.
It should be noted, however, that SLH-DSA remains viable for low-frequency, high-security operations such as firmware signing or root-of-trust authentication, where signing occurs rarely and the consistently low verification costs observed across all tested architectures make it suitable for endpoint deployment.
Hence, even more than the hardware, the architecture was found to have a major impact on signatures. A further finding was that whilst more capable machines such as the Pi 4 draw more power on-load (6.6 times more than the Pi 1), their much faster operations mean they consume less energy overall than the slower Pis.
\subsection{Limitations}
\textcolor{black}{Several} limitations \textcolor{black}{were identified} during the course of testing \textcolor{black}{that would be addressed in further work.} 

\textcolor{black}{The power and timed readings relied on averages,} which gave \textcolor{black}{the} typical value of the dataset; however, using the standard deviation could add a measure of how spread out the results are around that mean. Also, removing the 10\% top and low outlying data can be implemented to avoid spikes and anomalies to deliver as accurate data as possible.

Secondly, using the FNB58, \textcolor{black}{only PBUS result data was used to draw the conclusions.} Using the integrated Wh counter, \textcolor{black}{far more precise readings on actual usage could be obtained}, such as how many Joules are needed to generate 1000 Keys on a Pi Zero, instead of trying to measure the difference between on-load and baseline. 

\subsection{Future Works}
In future works around the viability of PQC, a next step would be to explore the ML-KEMs and ML-DSAs on even more constrained hardware, such as the Raspberry Pi Pico, which would require the engineering of a specific implementation since it cannot run OpenSSL.
Another avenue of research would be to explore actual IoT devices rather than Raspberry Pis: smart watches, FireTV sticks, old Android phones or TV boxes, IP cameras or infant monitors. Whilst this may appear beyond the scope of embedded cryptography research, it would match the sort of variety of architectures that may exist in industrial or medical contexts, where secure communications are critical and protection against harvest-now, decrypt-later is essential. 

\textcolor{black}{As mentioned in the limitations, the} Wh counter of the FNIRSI FNB58 \textcolor{black}{would be used in future work} to \textcolor{black}{obtain} more precise readings of power consumption. Access to a state-of-the-art oscilloscope could potentially give \textcolor{black}{access to} ultra-precise readings on individual operations, which may reinforce the findings.

Another further research question concerns the breadth of the signature landscape itself. This study restricted its signature analysis to the two standardised families, the lattice-based ML-DSA and the hash-based SLH-DSA, and to pure, single-algorithm configurations. Two extensions follow naturally. First, because the transition period is likely to see widespread deployment of hybrid constructions that pair a classical primitive with a post-quantum one, a comparison of pure versus hybrid signing and key-exchange on the same constrained hardware would quantify the cost of that conservatism across energy, latency, memory footprint and bandwidth: does the redundancy of a hybrid scheme impose a meaningful penalty on a 32-bit device? Second, NIST's additional digital-signature on-ramp has advanced candidates built on markedly different mathematical foundations, including FAEST, HAWK, MAYO, MQOM, QR-UOV, SDitH, SNOVA, SQIsign and UOV, spanning MPC-in-the-head, lattice, multivariate and isogeny constructions. These families differ not only in signature and key sizes but in their key-generation, signing and, crucially, signing-versus-verification cost asymmetries, which cannot be assumed to resemble those of the lattice-based and hash-based standards. A comparative study across this set would therefore address whether any emerging scheme is better matched to constrained or 32-bit IoT hardware than the current standards, and which of them retains the cheap-verification property that makes deployment on low-power endpoints attractive.

Finally, broader direction reframes the problem from a single device to a cloud-edge-IoT continuum, and from a benchmarking exercise into a systems-architecture question: how should cryptographic responsibilities be partitioned across cloud, edge and endpoint layers while preserving device identity, key protection and cryptographic agility? The asymmetry observed in this work, where verification was substantially cheaper than signing for the evaluated ML-DSA and SLH-DSA parameter sets (an asymmetry that cannot be assumed to hold for all signature families), suggests that constrained endpoints might perform only the comparatively less expensive operations while heavier work is delegated upward. Such delegation, however, is not purely a performance decision: offloading signing or key generation raises the question of where the private key resides, and many IoT trust models require an endpoint to hold its own identity key and sign locally. A first research question is therefore which operations can be safely delegated to edge or cloud nodes without violating device identity, trust or key-protection requirements. A second is one of cryptographic agility: as standards continue to evolve, with HQC, the additional-signature on-ramp and successive parameter revisions, how readily can a cloud-edge-IoT architecture migrate algorithms across all three tiers, and where do constrained endpoints become the bottleneck to that migration? These questions could be grounded empirically through a comparative study of managed cryptographic services across the major cloud providers, namely AWS, Google Cloud and Azure, measuring per-operation energy, cost and throughput and contrasting their ARM-based and x86-based instances, so as to extend the 32-versus-64-bit and ARM-versus-x86 findings of this paper from the embedded tier up to the datacentre.

\section{Supplementary}
The following supporting information can be downloaded at: \url{https://github.com/billbuchanan/pqc_energy}, accessed on June 2026: Python benchmark scripts (pqc\_benchmark.py, pqc\_benchmark\_automated.py), raw .cfn and .csv power reading files, and OpenSSL 3.5 build instructions.

Data are available in the supplementary materials and 
on GitHub at \url{https://github.com/billbuchanan/pqc_energy}, accessed on June 2026.

The authors wish to thank Andrey Volkov for the 
usb-meter-utils conversion utility.

\appendix
\section*{Power Readings and Detailed Results}

\begin{table}
\centering
\scriptsize
\caption{FNB58 average power readings in Watts across Raspberry Pi models}
\label{tab:power-readings}
\begin{longtable}{lccccccc}

\hline
 & Pi zero & Pi zero 2 & Pi 1 & Pi 2 1.1 & Pi 2 1.2 & Pi 3 & Pi 4 \\
\hline

Baseline & 0.5638 & 0.6451 & 2.2873 & 1.7490 & 1.7382 & 0.9592 & 1.5706 \\

\hline
\multicolumn{8}{l}{\textbf{Key generation}} \\

RSA-2048 & 0.9475 & 1.1074 & 2.4858 & 2.1170 & 2.0870 & 1.8235 & 2.5783 \\
ECDSA P-256 & 0.9013 & 1.2514 & 2.4986 & 2.1694 & 2.2446 & 1.9868 & 2.8906 \\
X25519 & 0.9096 & 1.2753 & 2.4997 & 2.1659 & 2.2503 & 1.9644 & 2.9676 \\
ECDH P-256 & 0.9064 & 1.3255 & 2.5011 & 2.1599 & 2.2700 & 1.9531 & 2.8855 \\
ML-KEM-512 & 0.9102 & 1.3219 & 2.5101 & 2.1710 & 2.2691 & 1.9093 & 2.9648 \\
ML-KEM-768 & 0.9157 & 1.2938 & 2.5028 & 2.1884 & 2.2686 & 1.9153 & 2.9834 \\
ML-KEM-1024 & 0.9136 & 1.3007 & 2.5259 & 2.1604 & 2.2525 & 1.9225 & 2.9777 \\
ML-DSA-44 & 0.9019 & 1.2714 & 2.5080 & 2.1665 & 2.2618 & 1.9145 & 2.9314 \\
ML-DSA-65 & 0.9171 & 1.2829 & 2.5050 & 2.2254 & 2.2733 & 1.9342 & 3.0068 \\
ML-DSA-87 & 0.9046 & 1.2858 & 2.5088 & 2.1589 & 2.2646 & 1.9359 & 2.9242 \\
SLH-DSA-SHA2-128s & 0.8988 & 1.3265 & 2.4817 & 2.0828 & 2.2725 & 1.9152 & 2.9935 \\
SLH-DSA-SHA2-128f & 0.9052 & 1.3017 & 2.4980 & 2.1500 & 2.2644 & 1.9153 & 2.9401 \\
SLH-DSA-SHAKE-128s & 0.9410 & 1.2568 & 2.4831 & 2.1033 & 2.2233 & 1.9080 & 2.8038 \\
SLH-DSA-SHA2-192s & 0.9176 & 1.3331 & 2.4768 & 2.0826 & 2.2763 & 1.9414 & 2.9744 \\
SLH-DSA-SHA2-192f & 0.9015 & 1.2858 & 2.4958 & 2.1395 & 2.2671 & 1.9324 & 2.9594 \\
SLH-DSA-SHAKE-192s & 0.9426 & 1.2693 & 2.4825 & 2.1015 & 2.2241 & 1.9301 & 2.8161 \\
SLH-DSA-SHA2-256s & 0.9024 & 1.3304 & 2.4713 & 2.0862 & 2.2780 & 1.9450 & 3.0044 \\
SLH-DSA-SHA2-256f & 0.9100 & 1.3067 & 2.4884 & 2.1248 & 2.2763 & 1.9594 & 3.0189 \\
SLH-DSA-SHAKE-256s & 0.9355 & 1.2653 & 2.4825 & 2.0989 & 2.2278 & 1.9495 & 2.8711 \\

\hline
\multicolumn{8}{l}{\textbf{Signatures - Signing}} \\

RSA-2048 & 0.9159 & 1.2496 & 2.4991 & 2.1701 & 2.2118 & 1.9317 & 2.8780 \\
ECDSA P-256 & 0.9035 & 1.2939 & 2.5049 & 2.1757 & 2.2780 & 1.9897 & 3.0196 \\
ML-DSA-44 & 0.9191 & 1.2603 & 2.5093 & 2.1641 & 2.2627 & 1.9858 & 2.8545 \\
ML-DSA-65 & 0.9060 & 1.2583 & 2.5059 & 2.1516 & 2.2410 & 1.9243 & 2.9670 \\
ML-DSA-87 & 0.9184 & 1.2492 & 2.5158 & 2.1395 & 2.2286 & 1.9441 & 2.9321 \\
SLH-DSA-SHA2-128s & 0.9027 & 1.3312 & 2.4760 & 2.0770 & 2.2812 & 1.9409 & 2.9777 \\
SLH-DSA-SHA2-128f & 0.9034 & 1.3307 & 2.4822 & 2.0868 & 2.2810 & 1.9255 & 2.9724 \\
SLH-DSA-SHAKE-128s & 0.9558 & 1.2763 & 2.4919 & 2.0974 & 2.2230 & 1.9218 & 2.8303 \\
SLH-DSA-SHA2-192s & 0.9032 & 1.3479 & 2.4699 & 2.0800 & 2.2797 & 1.9253 & 3.0154 \\
SLH-DSA-SHA2-192f & 0.9076 & 1.3190 & 2.4765 & 2.0831 & 2.2774 & 1.9507 & 2.9516 \\
SLH-DSA-SHAKE-192s & 0.9383 & 1.2665 & 2.4837 & 2.0898 & 2.2243 & 1.9436 & 2.8117 \\
SLH-DSA-SHA2-256s & 0.9072 & 1.3373 & 2.4728 & 2.0754 & 2.2798 & 1.9704 & 3.0005 \\
SLH-DSA-SHA2-256f & 0.9143 & 1.4109 & 2.4749 & 2.0802 & 2.2815 & 1.9412 & 2.9627 \\
SLH-DSA-SHAKE-256s & 0.9402 & 1.2814 & 2.4866 & 2.0894 & 2.2255 & 1.9317 & 2.8072 \\

\hline
\multicolumn{8}{l}{\textbf{Signatures - Verification}} \\

RSA-2048 & 0.8844 & 1.2424 & 2.4910 & 2.1477 & 2.2447 & 1.9633 & 2.7396 \\
ECDSA P-256 & 0.8932 & 1.2398 & 2.4956 & 2.1386 & 2.2496 & 1.9345 & 2.7338 \\
ML-DSA-44 & 0.8812 & 1.2506 & 2.4935 & 2.1498 & 2.2641 & 1.9497 & 2.7280 \\
ML-DSA-65 & 0.8877 & 1.2458 & 2.4998 & 2.1419 & 2.2570 & 1.9593 & 2.7572 \\
ML-DSA-87 & 0.8950 & 1.2572 & 2.5059 & 2.1437 & 2.2396 & 1.9385 & 2.7394 \\
SLH-DSA-SHA2-128s & 0.8839 & 1.2557 & 2.4959 & 2.1484 & 2.2536 & 1.9372 & 2.7623 \\
SLH-DSA-SHA2-128f & 0.8811 & 1.2803 & 2.4946 & 2.1161 & 2.2629 & 1.9433 & 2.8013 \\
SLH-DSA-SHAKE-128s & 0.8958 & 1.2501 & 2.4978 & 2.1282 & 2.2594 & 1.9630 & 2.7784 \\
SLH-DSA-SHA2-192s & 0.8861 & 1.2810 & 2.4985 & 2.1235 & 2.2635 & 1.9620 & 2.8333 \\
SLH-DSA-SHA2-192f & 0.8948 & 1.2761 & 2.4935 & 2.1112 & 2.2702 & 1.9582 & 2.8332 \\
SLH-DSA-SHAKE-192s & 0.9078 & 1.2552 & 2.4973 & 2.1368 & 2.2579 & 1.9421 & 2.7552 \\
SLH-DSA-SHA2-256s & 0.8893 & 1.2829 & 2.5000 & 2.1353 & 2.2565 & 1.9477 & 2.8197 \\
SLH-DSA-SHA2-256f & 0.9243 & 1.2735 & 2.4961 & 2.1219 & 2.2603 & 1.9557 & 2.8592 \\
SLH-DSA-SHAKE-256s & 0.9081 & 1.2795 & 2.5000 & 2.1347 & 2.2556 & 1.9182 & 2.7490 \\

\hline

\end{longtable}

\end{table}

\begin{longtable}{llccc}
\caption{Cryptographic Performance on Raspberry Pi Zero} \\
\hline
Algorithm & Operation & Time/Iter (s) & J/Iter & Iter/J \\
\hline
\endfirsthead

\hline
Algorithm & Operation & Time/Iter (s) & J/Iter & Iter/J \\
\hline
\endhead

\hline
\endfoot

\hline
\endlastfoot

\multicolumn{5}{l}{\textbf{Key Generation}} \\

RSA-2048 & KeyGen & 4.0040 & 1.5364 & 0.6509 \\
ECDSA P-256 & KeyGen & 0.1456 & 0.0491 & 20.3537 \\
X25519 & KeyGen & 0.1446 & 0.0500 & 20.0023 \\
ECDH P-256 & KeyGen & 0.1459 & 0.0500 & 20.0049 \\
ML-KEM-512 & KeyGen & 0.1492 & 0.0517 & 19.3462 \\
ML-DSA-44 & KeyGen & 0.1536 & 0.0519 & 19.2577 \\
SLH-DSA-SHA2-128f & KeyGen & 0.1887 & 0.0644 & 15.5202 \\
SLH-DSA-SHA2-128s & KeyGen & 2.7403 & 0.9182 & 1.0890 \\
SLH-DSA-SHAKE-192s & KeyGen & 6.8872 & 2.6092 & 0.3833 \\

\multicolumn{5}{l}{\textbf{Signing}} \\

RSA-2048 & Sign & 0.1360 & 0.0479 & 20.8875 \\
ECDSA P-256 & Sign & 0.0760 & 0.0258 & 38.7387 \\
ML-DSA-44 & Sign & 0.0899 & 0.0320 & 31.2894 \\
SLH-DSA-SHA2-128f & Sign & 1.0298 & 0.3776 & 2.6485 \\
SLH-DSA-SHA2-128s & Sign & 19.7340 & 7.2222 & 0.1385 \\
SLH-DSA-SHAKE-192s & Sign & 61.6652 & 24.7630 & 0.0404 \\

\multicolumn{5}{l}{\textbf{Verification}} \\

RSA-2048 & Verify & 0.0595 & 0.0207 & 48.3240 \\
ECDSA P-256 & Verify & 0.0618 & 0.0220 & 45.3843 \\
ML-DSA-44 & Verify & 0.0634 & 0.0218 & 45.7896 \\
SLH-DSA-SHA2-128f & Verify & 0.1188 & 0.0409 & 24.4461 \\
SLH-DSA-SHA2-128s & Verify & 0.0802 & 0.0278 & 35.9228 \\
SLH-DSA-SHAKE-192s & Verify & 0.1125 & 0.0417 & 23.9628 \\

\end{longtable}

\begin{longtable}{llccc}
\caption{Cryptographic Performance on Raspberry Pi Zero 2} \\
\hline
Algorithm & Operation & Time/Iter (s) & J/Iter & Iter/J \\
\hline
\endfirsthead

\hline
Algorithm & Operation & Time/Iter (s) & J/Iter & Iter/J \\
\hline
\endhead

\hline
\endfoot

\hline
\endlastfoot

\multicolumn{5}{l}{\textbf{Key Generation}} \\

RSA-2048 & KeyGen & 0.7825 & 0.3617 & 2.7645 \\
ECDSA P-256 & KeyGen & 0.0564 & 0.0342 & 29.2711 \\
X25519 & KeyGen & 0.0553 & 0.0348 & 28.7125 \\
ECDH P-256 & KeyGen & 0.0561 & 0.0382 & 26.1859 \\
ML-KEM-512 & KeyGen & 0.0574 & 0.0388 & 25.7429 \\
ML-DSA-44 & KeyGen & 0.0592 & 0.0371 & 26.9736 \\
SLH-DSA-SHA2-128f & KeyGen & 0.0706 & 0.0464 & 21.5619 \\
SLH-DSA-SHA2-128s & KeyGen & 0.9195 & 0.6265 & 1.5961 \\
SLH-DSA-SHAKE-192s & KeyGen & 1.2198 & 0.7613 & 1.3135 \\

\multicolumn{5}{l}{\textbf{Signing}} \\

RSA-2048 & Sign & 0.0426 & 0.0257 & 38.8494 \\
ECDSA P-256 & Sign & 0.0296 & 0.0192 & 52.1412 \\
ML-DSA-44 & Sign & 0.0348 & 0.0214 & 46.6919 \\
SLH-DSA-SHA2-128f & Sign & 0.3441 & 0.2359 & 4.2387 \\
SLH-DSA-SHA2-128s & Sign & 6.5766 & 4.5117 & 0.2216 \\
SLH-DSA-SHAKE-192s & Sign & 10.5073 & 6.5288 & 0.1532 \\

\multicolumn{5}{l}{\textbf{Verification}} \\

RSA-2048 & Verify & 0.0229 & 0.0137 & 73.0033 \\
ECDSA P-256 & Verify & 0.0238 & 0.0142 & 70.6107 \\
ML-DSA-44 & Verify & 0.0249 & 0.0151 & 66.3558 \\
SLH-DSA-SHA2-128f & Verify & 0.0429 & 0.0272 & 36.7106 \\
SLH-DSA-SHA2-128s & Verify & 0.0301 & 0.0184 & 54.4766 \\
SLH-DSA-SHAKE-192s & Verify & 0.0325 & 0.0198 & 50.4402 \\

\end{longtable}

\begin{longtable}{llccc}
\caption{Cryptographic Performance on Raspberry Pi 1} \\
\hline
Algorithm & Operation & Time/Iter (s) & J/Iter & Iter/J \\
\hline
\endfirsthead

\hline
Algorithm & Operation & Time/Iter (s) & J/Iter & Iter/J \\
\hline
\endhead

\hline
\endfoot

\hline
\endlastfoot

\multicolumn{5}{l}{\textbf{Key Generation}} \\

RSA-2048 & KeyGen & 6.3753 & 1.2655 & 0.7902 \\
ECDSA P-256 & KeyGen & 0.2252 & 0.0476 & 21.0158 \\
X25519 & KeyGen & 0.2263 & 0.0481 & 20.8074 \\
ECDH P-256 & KeyGen & 0.2257 & 0.0483 & 20.7230 \\
ML-KEM-512 & KeyGen & 0.2328 & 0.0519 & 19.2843 \\
ML-DSA-44 & KeyGen & 0.2379 & 0.0525 & 19.0437 \\
SLH-DSA-SHA2-128f & KeyGen & 0.2854 & 0.0601 & 16.6319 \\
SLH-DSA-SHA2-128s & KeyGen & 3.7332 & 0.7259 & 1.3777 \\
SLH-DSA-SHAKE-192s & KeyGen & 10.5211 & 2.0534 & 0.4870 \\

\multicolumn{5}{l}{\textbf{Signing}} \\

RSA-2048 & Sign & 0.2059 & 0.0436 & 22.9370 \\
ECDSA P-256 & Sign & 0.1167 & 0.0254 & 39.3977 \\
ML-DSA-44 & Sign & 0.1366 & 0.0303 & 32.9924 \\
SLH-DSA-SHA2-128f & Sign & 1.4149 & 0.2757 & 3.6269 \\
SLH-DSA-SHA2-128s & Sign & 27.2156 & 5.1354 & 0.1947 \\
SLH-DSA-SHAKE-192s & Sign & 95.5478 & 18.7602 & 0.0533 \\

\multicolumn{5}{l}{\textbf{Verification}} \\

RSA-2048 & Verify & 0.0915 & 0.0186 & 53.6424 \\
ECDSA P-256 & Verify & 0.0945 & 0.0197 & 50.7856 \\
ML-DSA-44 & Verify & 0.0969 & 0.0200 & 50.0614 \\
SLH-DSA-SHA2-128f & Verify & 0.1727 & 0.0358 & 27.9336 \\
SLH-DSA-SHA2-128s & Verify & 0.1198 & 0.0250 & 40.0089 \\
SLH-DSA-SHAKE-192s & Verify & 0.1716 & 0.0360 & 27.7494 \\

\end{longtable}

\begin{longtable}{llccc}
\caption{Cryptographic Performance on Raspberry Pi 2 Model B v1.1 (32-bit)} \\
\hline
Algorithm & Operation & Time/Iter (s) & J/Iter & Iter/J \\
\hline
\endfirsthead

\hline
Algorithm & Operation & Time/Iter (s) & J/Iter & Iter/J \\
\hline
\endhead

\hline
\endfoot

\hline
\endlastfoot

\multicolumn{5}{l}{\textbf{Key Generation}} \\

RSA-2048 & KeyGen & 3.4957 & 1.2864 & 0.7773 \\
ECDSA P-256 & KeyGen & 0.0938 & 0.0393 & 25.4235 \\
X25519 & KeyGen & 0.0911 & 0.0380 & 26.3417 \\
ECDH P-256 & KeyGen & 0.0925 & 0.0380 & 26.3194 \\
ML-KEM-512 & KeyGen & 0.0942 & 0.0398 & 25.1484 \\
ML-DSA-44 & KeyGen & 0.0986 & 0.0412 & 24.2826 \\
SLH-DSA-SHA2-128f & KeyGen & 0.1253 & 0.0503 & 19.8998 \\
SLH-DSA-SHA2-128s & KeyGen & 2.1249 & 0.7093 & 1.4098 \\
SLH-DSA-SHAKE-192s & KeyGen & 6.5216 & 2.2992 & 0.4349 \\

\multicolumn{5}{l}{\textbf{Signing}} \\

RSA-2048 & Sign & 0.1023 & 0.0431 & 23.2172 \\
ECDSA P-256 & Sign & 0.0483 & 0.0206 & 48.5266 \\
ML-DSA-44 & Sign & 0.0605 & 0.0251 & 39.7828 \\
SLH-DSA-SHA2-128f & Sign & 0.8025 & 0.2711 & 3.6888 \\
SLH-DSA-SHA2-128s & Sign & 15.6621 & 5.1381 & 0.1946 \\
SLH-DSA-SHAKE-192s & Sign & 57.6643 & 19.6558 & 0.0509 \\

\multicolumn{5}{l}{\textbf{Verification}} \\

RSA-2048 & Verify & 0.0394 & 0.0157 & 63.5858 \\
ECDSA P-256 & Verify & 0.0429 & 0.0167 & 59.8806 \\
ML-DSA-44 & Verify & 0.0434 & 0.0174 & 57.4843 \\
SLH-DSA-SHA2-128f & Verify & 0.0842 & 0.0309 & 32.3551 \\
SLH-DSA-SHA2-128s & Verify & 0.0549 & 0.0219 & 45.5920 \\
SLH-DSA-SHAKE-192s & Verify & 0.0903 & 0.0350 & 28.5602 \\

\end{longtable}

\begin{longtable}{llccc}
\caption{Cryptographic Performance on Raspberry Pi 2 Model B v1.2 (64-bit)} \\
\hline
Algorithm & Operation & Time/Iter (s) & J/Iter & Iter/J \\
\hline
\endfirsthead

\hline
Algorithm & Operation & Time/Iter (s) & J/Iter & Iter/J \\
\hline
\endhead

\hline
\endfoot

\hline
\endlastfoot

\multicolumn{5}{l}{\textbf{Key Generation}} \\

RSA-2048 & KeyGen & 0.9099 & 0.3173 & 3.1513 \\
ECDSA P-256 & KeyGen & 0.0655 & 0.0332 & 30.1633 \\
X25519 & KeyGen & 0.0644 & 0.0330 & 30.3016 \\
ECDH P-256 & KeyGen & 0.0650 & 0.0346 & 28.9230 \\
ML-KEM-512 & KeyGen & 0.0672 & 0.0357 & 28.0464 \\
ML-DSA-44 & KeyGen & 0.0689 & 0.0361 & 27.7302 \\
SLH-DSA-SHA2-128f & KeyGen & 0.0820 & 0.0431 & 23.1852 \\
SLH-DSA-SHA2-128s & KeyGen & 1.0335 & 0.5521 & 1.8111 \\
SLH-DSA-SHAKE-192s & KeyGen & 1.3695 & 0.6654 & 1.5028 \\

\multicolumn{5}{l}{\textbf{Signing}} \\

RSA-2048 & Sign & 0.0491 & 0.0232 & 43.0179 \\
ECDSA P-256 & Sign & 0.0341 & 0.0184 & 54.3523 \\
ML-DSA-44 & Sign & 0.0416 & 0.0218 & 45.8454 \\
SLH-DSA-SHA2-128f & Sign & 0.3875 & 0.2103 & 4.7543 \\
SLH-DSA-SHA2-128s & Sign & 7.3566 & 3.9943 & 0.2504 \\
SLH-DSA-SHAKE-192s & Sign & 11.7895 & 5.7308 & 0.1745 \\

\multicolumn{5}{l}{\textbf{Verification}} \\

RSA-2048 & Verify & 0.0267 & 0.0135 & 73.9650 \\
ECDSA P-256 & Verify & 0.0275 & 0.0141 & 71.0128 \\
ML-DSA-44 & Verify & 0.0289 & 0.0152 & 65.8137 \\
SLH-DSA-SHA2-128f & Verify & 0.0482 & 0.0253 & 39.5586 \\
SLH-DSA-SHA2-128s & Verify & 0.0381 & 0.0196 & 50.9929 \\
SLH-DSA-SHAKE-192s & Verify & 0.0373 & 0.0194 & 51.5813 \\

\end{longtable}

\begin{longtable}{llccc}
\caption{Cryptographic Performance on Raspberry Pi 3 Model A+} \\
\hline
Algorithm & Operation & Time/Iter (s) & J/Iter & Iter/J \\
\hline
\endfirsthead

\hline
Algorithm & Operation & Time/Iter (s) & J/Iter & Iter/J \\
\hline
\endhead

\hline
\endfoot

\hline
\endlastfoot

\multicolumn{5}{l}{\textbf{Key Generation}} \\

RSA-2048 & KeyGen & 0.5702 & 0.4928 & 2.0292 \\
ECDSA P-256 & KeyGen & 0.0442 & 0.0454 & 22.0284 \\
X25519 & KeyGen & 0.0432 & 0.0435 & 23.0061 \\
ECDH P-256 & KeyGen & 0.0439 & 0.0436 & 22.9434 \\
ML-KEM-512 & KeyGen & 0.0450 & 0.0428 & 23.3784 \\
ML-DSA-44 & KeyGen & 0.0464 & 0.0444 & 22.5447 \\
SLH-DSA-SHA2-128f & KeyGen & 0.0549 & 0.0525 & 19.0351 \\
SLH-DSA-SHA2-128s & KeyGen & 0.6594 & 0.6304 & 1.5862 \\
SLH-DSA-SHAKE-192s & KeyGen & 0.8739 & 0.8485 & 1.1786 \\

\multicolumn{5}{l}{\textbf{Signing}} \\

RSA-2048 & Sign & 0.0326 & 0.0317 & 31.5857 \\
ECDSA P-256 & Sign & 0.0234 & 0.0241 & 41.4915 \\
ML-DSA-44 & Sign & 0.0277 & 0.0284 & 35.2079 \\
SLH-DSA-SHA2-128f & Sign & 0.2480 & 0.2396 & 4.1731 \\
SLH-DSA-SHA2-128s & Sign & 4.6841 & 4.5984 & 0.2175 \\
SLH-DSA-SHAKE-192s & Sign & 7.4921 & 7.3749 & 0.1356 \\

\multicolumn{5}{l}{\textbf{Verification}} \\

RSA-2048 & Verify & 0.0186 & 0.0187 & 53.5543 \\
ECDSA P-256 & Verify & 0.0189 & 0.0184 & 54.3356 \\
ML-DSA-44 & Verify & 0.0200 & 0.0198 & 50.4455 \\
SLH-DSA-SHA2-128f & Verify & 0.0324 & 0.0319 & 31.3875 \\
SLH-DSA-SHA2-128s & Verify & 0.0238 & 0.0232 & 43.0466 \\
SLH-DSA-SHAKE-192s & Verify & 0.0246 & 0.0242 & 41.3491 \\

\end{longtable}

\begin{longtable}{llccc}
\caption{Cryptographic Performance on Raspberry Pi 4} \\
\hline
Algorithm & Operation & Time/Iter (s) & J/Iter & Iter/J \\
\hline
\endfirsthead

\hline
Algorithm & Operation & Time/Iter (s) & J/Iter & Iter/J \\
\hline
\endhead

\hline
\endfoot

\hline
\endlastfoot

\multicolumn{5}{l}{\textbf{Key Generation}} \\

RSA-2048 & KeyGen & 0.3526 & 0.3553 & 2.8141 \\
ECDSA P-256 & KeyGen & 0.0247 & 0.0326 & 30.7203 \\
X25519 & KeyGen & 0.0239 & 0.0334 & 29.9306 \\
ECDH P-256 & KeyGen & 0.0243 & 0.0319 & 31.3277 \\
ML-KEM-512 & KeyGen & 0.0248 & 0.0345 & 28.9628 \\
ML-DSA-44 & KeyGen & 0.0255 & 0.0347 & 28.8221 \\
SLH-DSA-SHA2-128f & KeyGen & 0.0297 & 0.0407 & 24.5891 \\
SLH-DSA-SHA2-128s & KeyGen & 0.3241 & 0.4612 & 2.1682 \\
SLH-DSA-SHAKE-192s & KeyGen & 0.5114 & 0.6370 & 1.5699 \\

\multicolumn{5}{l}{\textbf{Signing}} \\

RSA-2048 & Sign & 0.0194 & 0.0254 & 39.3510 \\
ECDSA P-256 & Sign & 0.0136 & 0.0197 & 50.6943 \\
ML-DSA-44 & Sign & 0.0156 & 0.0200 & 50.0352 \\
SLH-DSA-SHA2-128f & Sign & 0.1234 & 0.1730 & 5.7811 \\
SLH-DSA-SHA2-128s & Sign & 2.2925 & 3.2258 & 0.3100 \\
SLH-DSA-SHAKE-192s & Sign & 4.4086 & 5.4714 & 0.1828 \\

\multicolumn{5}{l}{\textbf{Verification}} \\

RSA-2048 & Verify & 0.0101 & 0.0118 & 85.0453 \\
ECDSA P-256 & Verify & 0.0112 & 0.0130 & 76.8576 \\
ML-DSA-44 & Verify & 0.0124 & 0.0143 & 69.8455 \\
SLH-DSA-SHA2-128f & Verify & 0.0170 & 0.0210 & 47.6736 \\
SLH-DSA-SHA2-128s & Verify & 0.0125 & 0.0149 & 67.0421 \\
SLH-DSA-SHAKE-192s & Verify & 0.0138 & 0.0164 & 61.0741 \\

\end{longtable}

\bibliographystyle{IEEEtran}
\bibliography{references}

\end{document}